\documentclass[aps,pra,preprint,superscriptaddress,amsmath,amssymb,showkeys]{revtex4}
\usepackage{graphicx}
\begin{document}

\title{On axoplasmic pressure waves and their possible role in nerve impulse propagation}

\author{Marat M. Rvachev}

\affiliation{rvachev@alum.mit.edu}

\date{June 7, 2010}

\begin{abstract}
It is suggested that the propagation of the action potential is accompanied by an axoplasmic pressure pulse propagating in the axoplasm along the axon length. The pressure pulse stretch-modulates voltage-gated Na$^+$ (Nav) channels embedded in the axon membrane, causing their accelerated activation and inactivation and increasing peak channel conductance. As a result, the action potential propagates due to mechano-electrical activation of Nav channels by straggling ionic currents and the axoplasmic pressure pulse. The velocity of such propagation is higher than in the classical purely electrical Nav activation mechanism, and it may be close to the velocity of propagation of pressure pulses in the axoplasm. Extracellular Ca$^{2+}$ ions influxing during the voltage spike, or Ca$^{2+}$ ions released from intracellular stores, may trigger a mechanism that generates and augments the pressure pulse, thus opposing its viscous decay. The model can potentially explain a number of phenomena that are not contained within the purely electrical Hodgkin-Huxley-type framework: the Meyer-Overton rule for the effectiveness of anesthetics, as well as various mechanical, optical and thermodynamic phenomena accompanying the action potential. It is shown that the velocity of propagation of axoplasmic pressure pulses is close to the measured velocity of the nerve impulse, both in absolute magnitude and in dependence on axon diameter, degree of myelination and temperature.

\keywords{Axoplasmic pressure waves; action potential; Hodgkin-Huxley model; Meyer-Overton rule.}
\end{abstract}

\maketitle

\section{Introduction}	

The well-known Hodgkin-Huxley (H-H) model of propagation of the nerve impulse in neuronal axons \cite{hodg52} has recently been the subject of constructive criticism \cite{heim05,heim07}. As a purely electrical model, the H-H model does not contain a number of non-electrical phenomena which need to be included in a more complete model of axon function. These are changes in nerve dimensions and in the normal force exerted by the nerve \cite{iwas80b,iwas80,tasa80,tasa82,tasa90}, reversible changes in temperature and heat \cite{abbo58,howa68,ritc85,tasa89,tasa92} and changes in fluorescence intensity and anisotropy of lipid membrane markers \cite{tasa68,tasa69}, all observed synchronously with a propagating action potential. Interestingly, within the context of presently accepted models it has not been possible to explain the famous Meyer-Overton rule \cite{meye99,over01} for the effectiveness of anesthetics on nerve fibers. The rule states that the effectiveness of an anesthetic is linearly related to the solubility of that anesthetic in membranes \cite{urba06}. The rule is valid over five orders of magnitude and holds independently of the chemical identity of the anesthetic; it is valid for noble gases such as argon and xenon as well as for alkanols. Although some ion channels are influenced by some anesthetics, there is no quantitative correlation with the Meyer-Overton rule \cite{heim07}, which makes it difficult to interpret the rule within the framework of the H-H type excitation kinetics. Although H-H type models have been very successful at reconstructing various aspects of action potential propagation \cite{hodg52,cool66}, new mechanisms may need to be considered to explain the Meyer-Overton rule.

In this paper, we suggest an extension of the H-H model of the propagation of the action potential. We suggest that its propagation may be aided by stretch-modulation of membrane voltage-gated channels by an axoplasmic pressure pulse. This may lead to the action potential that propagates at a velocity close to the velocity of propagation of the axoplasmic pressure pulse. For both myelinated and unmyelinated fibers of various diameters, we estimate the pressure pulse velocity and indeed find that it agrees very well with experimental values for the velocity of the nerve impulse. The model offers explanation for the Meyer-Overton rule as well as for various mechanical, optical and thermodynamic phenomena that accompany the action potential. 
 The speculation that density pulses may be related to nerve impulses has been discussed by various authors since 1912  \cite{wilk12,wilk12b,hodg45,kauf89}.
Most recently, it has been suggested \cite{heim05,heim07} that the action potential is a mechanical soliton (density pulse) propagating in the axon membrane. While this model appears to account for the above-mentioned phenomena, as well as for the Meyer-Overton rule, it does not seem to explain changes in the propagation velocity that are related to axon diameter.

\section{The Pressure Wave Model}

We suggest that the propagation of the nerve impulse may involve an axoplasmic pressure pulse propagating roughly synchronously with the wave of membrane depolarization. While several variations of the process can be envisioned, we suggest that it proceeds as outlined in Fig. 1. 
\begin{figure}[tb]
\includegraphics[width=15cm]{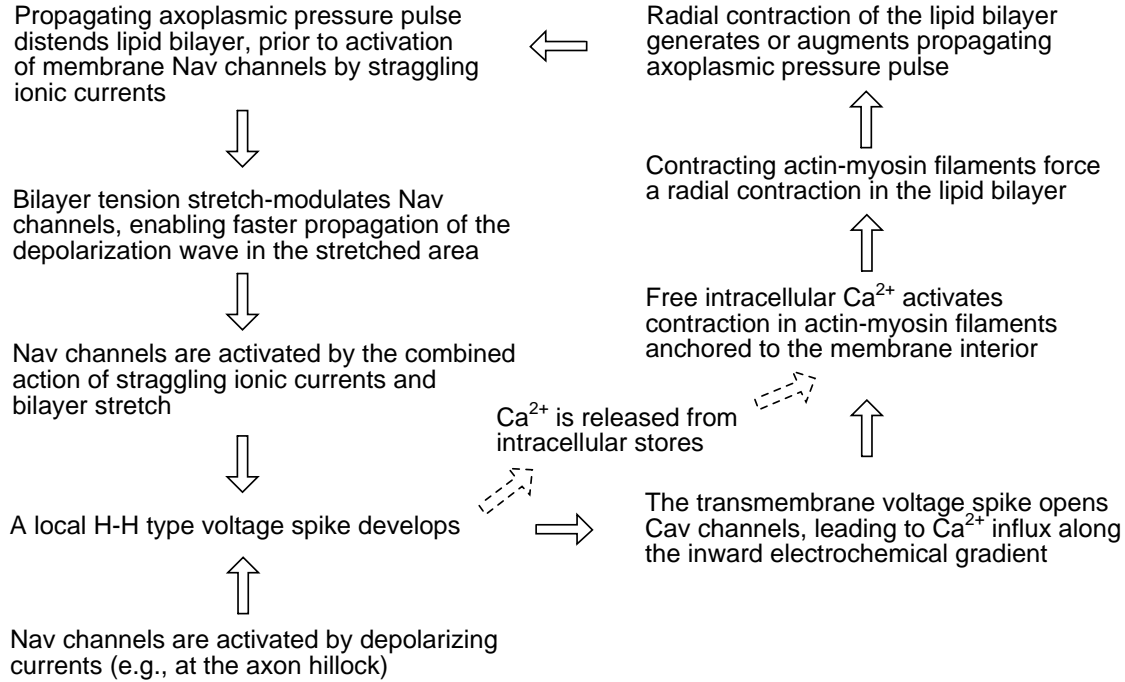}
\vspace*{8pt}
\caption{\label{fig2} The suggested process of propagation of the action potential and the axoplasmic pressure pulse.}
\end{figure}
The axoplasmic pressure pulse is initiated by an action potential as discussed later in this section. The propagating axoplasmic pressure pulse distends the axon membrane prior to activation of membrane voltage-gated Na$^+$ (Nav) channels by the straggling ionic currents. This can occur if the axoplasmic pressure pulse propagates faster than the action potential associated with the H-H process for unstretched membrane. The stretch in the lipid bilayer modulates Nav channels, allowing a faster propagating H-H type excitation wave in the stretched membrane by enabling accelerated channel activation and inactivation and increasing peak channel conductance, as well as possibly lowering channel activation threshold. If this faster propagating H-H excitation wave is at least as fast as the traveling pressure wave, the two waves will co-propagate. In such a case, activation and development of local voltage spikes will be driven by the combined action of the membrane depolarization by the straggling ionic currents and the membrane stretch caused by the propagating pressure pulse. Further, we suggest the following mechanism of generation and amplification of the axoplasmic pressure pulse. (Below it is shown that under physiological conditions axoplasmic pressure pulses decay over roughly 1 mm distance due to viscosity, and therefore sustained propagation in long axons requires a mechanism of amplification.) A local H-H  type voltage spike (which may be caused by an initial membrane depolarization such as at the axon hillock, or by the propagating action potential) activates membrane voltage-gated Ca$^{2+}$ (Cav) channels, leading to an influx of Ca$^{2+}$ ions along their inward electrochemical gradient; or, the voltage spike triggers Ca$^{2+}$ release from intracellular stores. The presence of free intracellular Ca$^{2+}$ induces contraction of actin-myosin filaments anchored to the membrane interior. The filament contraction forces a radial contraction in cylindrical segments of the membrane (Fig. 2), which generates or augments the propagating pressure pulse. 

\begin{figure}[t]
\includegraphics[height=4cm]{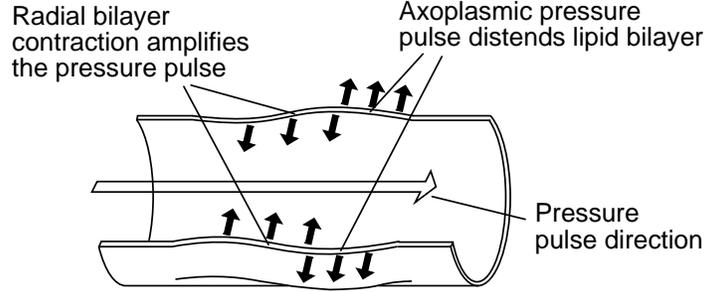}
\vspace*{8pt}
\caption{\label{fig:fig3} Amplification of the axoplasmic pressure pulse by a radial contraction  of the axon membrane (drawn not to scale).}
\end{figure}

Figure 3 illustrates suggested processes of mechanical action of the axoplasmic pressure pulse on Nav channels. 
\begin{figure}[t]
\includegraphics[width=9cm]{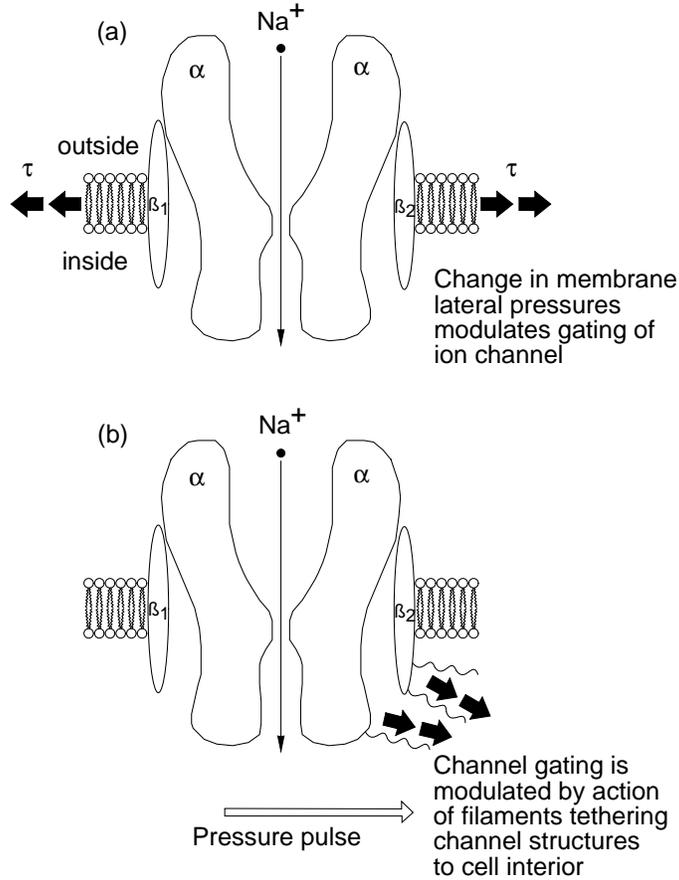}
\vspace*{8pt}
\caption{\label{fig4} Mechanical action of a traveling axoplasmic pressure pulse on Nav channels (see text).}
\end{figure}
The lipid bilayer is stretched laterally (Fig. 3(a)) as the propagating pressure pulse causes a $\Delta R(t)$ increase in the membrane cylinder radius $R$, corresponding to a $\Delta\tau(t)$ increase in the lipid bilayer tension $\tau$. Tension $\tau$ can be expressed through the bilayer lateral pressure profile $p(z)$ \cite{cant01}: $\tau=\int_{-\frac{h}{2}}^{\frac{h}{2}}p(z) dz$, where $z$ is the depth within the bilayer, and the integral is over the bilayer thickness $h$. The change $\Delta\tau(t)$ in the bilayer tension results in a change $\Delta p(z,t)$ in the lateral pressure profile such that $\Delta\tau(t)=\int_{-\frac{h}{2}}^{\frac{h}{2}}\Delta p(z,t) dz$. We suggest that these alterations in the bilayer lateral pressures modulate voltage-gating properties of Nav channels embedded in the bilayer. Alternatively, it is plausible that Nav channels are modulated by direct mechanical links (such as filaments) tethering the ion channel structures to the interior cytoskeleton that is perturbed by the propagating pressure pulse, as shown in Fig. 3(b). 

The model presented requires that Nav channels be mechanosensitive and allow a (perhaps substantially) faster propagating H-H excitation wave in a stretched membrane. Nav, Kv and Cav channels all exhibit reversible gating changes with stretch \cite{gu01,morr07}. In experiments with pipette pressure applied to a membrane patch \cite{morr07}, Nav channel activation and inactivation was reversibly accelerated by up to 1.85-fold (1.4-fold for typical stimuli used). Stretch also increased peak $I_{Na}$ by about 1.5-fold near the activation threshold for moderate stretch stimuli. It is clear that such significant changes in Nav gating should lead to a faster, and probably much faster, propagating H-H excitation. The above-mentioned experiments used stretch stimuli lasting on the order of tens of seconds. It is not excluded that for much shorter, $\sim$1~ms, stretch stimuli, Nav channels may also exhibit different gating changes, such as a change in the channel activation threshold. 

According to the model presented, the action potential propagates as a result of mechano-electrical activation of Nav channels by both straggling axoplasmic ionic currents and an axoplasmic pressure wave. Further, the transmembrane voltage spike elevates intracellular Ca$^{2+}$ concentration ([Ca$^{2+}$]$_\textnormal{i}$), either by activating membrane Cav channels and allowing entry of external Ca$^{2+}$ into the cell, or by Ca$^{2+}$ release from intracellular stores. Then, free intracellular Ca$^{2+}$ initiates a contraction in the filament network anchored to the membrane interior. Ca$^{2+}$ is a good candidate for transducing the H-H voltage spike into filament contraction for several reasons. First, Ca$^{2+}$ ion channels are ubiquitous. To quote from Ref.~\cite{hill01}, "voltage-gated Ca channels are found in every excitable cell. In fact, I feel they define excitable cells." Ca$^{2+}$ channels are often overlooked, as they are usually found in low density, with typical ionic fluxes during an action potential 100-fold lower than Na fluxes \cite{hill01}. Furthermore, many types of Ca$^{2+}$ ion channels are directly and rapidly gated by voltage \cite{nau08}. Also, an increase in [Ca$^{2+}$]$_\textnormal{i}$ is often associated with initiation of motion in cells, from motility in freely moving cells and muscle contraction to synaptic vesicle release at synapses. [Ca$^{2+}$]$_\textnormal{i}$ in a resting cell is kept extremely low (30-200 nM) by the combined actions of a Ca$^{2+}$ pump and a Na$^+$-Ca$^{2+}$ exchange system on the surface membrane and by Ca$^{2+}$ pumps on intracellular organelles; extracellular Ca$^{2+}$ concentration ([Ca$^{2+}$]$_\textnormal{o}$) is normally much higher. Ca$^{2+}$ ions that influx during the action potential should instantly increase the Ca$^{2+}$ level many fold in the vicinity of the filament network attached to the inner membrane surface; that, coupled with the ability of Ca$^{2+}$ to induce rapid conformational changes in proteins (such as upon its binding to actin filaments in the actin-myosin muscle complex), could provide a fast contractile response necessary to amplify a propagating pressure pulse. Experiments show a drastic change in nerve impulse conduction in low [Ca$^{2+}$]$_\textnormal{o}$, with the conduction poor or blocked at initiation of cathodal current; these nerve fibers do respond better to anodal break excitation, which provides a higher gradient for Ca$^{2+}$ entry \cite{hill77}. However, both [Ca$^{2+}$]$_\textnormal{i}$ and [Ca$^{2+}$]$_\textnormal{o}$ significantly affect gating of voltage-gated channels (e.g.,  [Ca$^{2+}$]$_\textnormal{o}$ meaningfully shifts gating thresholds of all voltage-gated channels) \cite{hill01}, which affects propagation in the pure H-H framework as well. This may be a reason why measurements of nerve impulse conduction velocity with depleted [Ca$^{2+}$]$_\textnormal{o}$ are not readily available in the literature. We note that, since the radial bilayer contraction follows the pressure pulse at the same velocity, even a small contraction that yields a small amplification will accumulate over distance and, thus, create a larger amplifying effect. Within this context, it should also be noted that a pressure pulse that propagates in an inviscid incompressible fluid that is enclosed in a viscoelastic tube will dissipate energy and subsequently decay. This decay occurs because the pressure exerted by the tube on the fluid during tube radial expansion is greater than that exerted during tube radial contraction. As a result, the total work done by the tube on the fluid is negative. In axons, the situation may be the reverse, in which a larger pressure exerted by the bilayer during its contraction may result in the overall transmission of kinetic energy to the axoplasm. We also posit that generation and amplification of the axoplasmic pressure pulse may proceed through electro-mechanical coupling such as voltage-induced membrane movement \cite{zhan01} resulting from the H-H voltage spike. Apart from membrane movement, it is plausible that elevated [Ca$^{2+}$]$_\textnormal{i}$ associated with a propagating action potential may act on interior actin-myosin and other cytoskeletal elements to cause local cytoplasmic pressure variations that, accumulated over distance, may lead to generation of the axoplasmic pressure pulse. 

\section{Propagation of Axoplasmic Pressure Pulses}

We consider the propagation of small amplitude, axially symmetric pressure pulses in a viscous compressible axoplasm enclosed in a circular cylindrical thin-walled distensible membrane. The theoretical analysis of waves in viscous fluids enclosed in tubes has been presented in Refs.~\cite{rayl45,morg54,rubi78}. For the limiting case of a viscous compressible axoplasm in a rigid tube, the pressure pulse is an axoplasmic density disturbance that propagates along the axon similarly to a sound wave packet, with the pulse potential energy stored in the axoplasm bulk deformation \cite{rayl45,rubi78}. In the opposing limit of a viscous incompressible axoplasm in a distensible membrane, the pressure pulse manifests itself as an increase in the axon diameter and associated membrane area expansion, with the pulse potential energy stored in the membrane strain \cite{morg54,rubi78}.
Let us consider a pressure pulse of central frequency $\omega$ propagating in the axoplasm of density $\rho$, compressibility $\kappa$ and dynamic viscosity $\mu$, enclosed in a cylindrical membrane of radius $R$, thickness $h$ and area expansion modulus $K$. Let us also assume that for the deformation of area expansion (when the membrane material behaves in an elastic solid manner), the membrane Young's modulus is $E$ and the Poisson's ratio is $\nu$.
If viscosity is neglected, the pulse propagates with the velocity \cite{chev93}:
\begin{equation}
v_0=\sqrt{\frac{1}{\rho(\kappa + \frac{2R}{K})}}.
\end{equation}
For typical unmyelinated axons, i.e., assuming $K=0.8$~N/m (twice the single plasma membrane area expansion modulus of 0.4 N/m \cite{thom97}, to account for the combination of the axon and adjoining glial cell membranes), 
$\kappa =4.04\cdot10^{-10}$ Pa$^{-1}$ (water at 38 $^\circ$C \cite{lide02}), $\rho = 1050$~kg/m$^3$ (axoplasm \cite{keyn51}), and $2R=1$ $\mu$m, Eq. (1) yields the pressure pulse velocity of $28$ m/s. 
Note that if the axon membrane is assumed to be indistensible ($\frac{2R}{K}\ll\kappa$), Eq. (1) yields a pulse velocity of 1535 m/s. 

\subsection{Viscous axoplasm}

Viscous forces in the axoplasm and in the axon membrane in general decrease the velocity of axoplasmic pressure waves and introduce a decay \cite{rayl45,morg54,rubi78}. The membrane viscous tension is characterized by the ratio of the surface viscosity (on the order of 10$^{-6}$ N$\cdot$s/m \cite{evan76}) and the time scale of the pressure pulse (here assumed to be close to the action potential duration, about 1 ms), yielding 10$^{-3}$ N/m. Since this value is much smaller than the membrane elastic area expansion modulus, 0.4 N/m, the effects of membrane viscosity on the pulse propagation can be neglected \cite{rubi78}. Concerning the viscosity of the axoplasm, for the general case of a viscous compressible fluid enclosed in an elastic tube, the dispersion equation is quite complicated and has not been solved in closed form. However, the equation is simplified and can be solved in the high viscosity limit, defined by \cite{rayl45,morg54,rubi78}:
\begin{equation}
\alpha \equiv R \sqrt{\frac{\omega\rho}{\mu}} \ll 1.
\end{equation}
Here we evaluate Eq. (2) for typical axons. As before, we assume that the duration of the axoplasmic pressure pulse is similar to that of the action potential, about 1 ms, yielding $\omega \approx 3142$ rad/s. Axoplasm viscosity $\mu$, as is relevant for perturbations associated with small-amplitude axoplasmic pressure pulses, is assumed to be similar to the viscosity of water, $\mu=6.82 \cdot 10^{-4}$ Pa$\cdot$s at 38 $^\circ$C \cite{lide02}. Taking $2R=10 \mu$m (the typical internal myelin diameter for A$\alpha$ myelinated fibers) and $\rho = 1050$~kg/m$^3$ yields $\alpha = 0.35$, which should be small enough for high-viscosity approximation to the dispersion equation to hold \cite{rubi78} for axons of similar or smaller diameters. When $\alpha\gg 1$ ($2R\gg29\mu$m), the effects of viscosity are small, and the pressure pulse propagates with little decay and with the velocity described by Eq. (1).

When Eq. (2) holds, the phase velocity of axoplasmic pressure waves (the velocity of the harmonic waves) is given by \cite{rayl45,morg54,rubi78}:
\begin{equation}
v_{ph}=\frac{c}{2} \cdot \alpha \cdot v_0,
\end{equation}
where $\alpha$ and $v_0$ are given by Eqs. (2) and (1), respectively; $c=1$ when $\frac{2R}{K} \ll \kappa$ (the limit of a rigid membrane), while $c=\frac{2}{\sqrt{5-4\nu}}$ when $\frac{2R}{K} \gg \kappa$ (the limit of a soft membrane). The factor $c$ accounts for movement of the membrane in the axial direction caused by the viscous axoplasm in the limit of a soft membrane; here and below we assume $\nu=0.5$ for the incompressible lipid bilayer, leading to $\frac{2}{\sqrt{5-4\nu}}\approx 1.15$.
The velocity of the pressure pulse (the "group" velocity) is given by $v_{gr}=\frac{\partial \omega}{\partial k}$,
where $k$ is the wavenumber.
From Eqs. (1-3), and using the relations $v_{ph}=\frac{\omega}{k}$ and $v_{gr}=\frac{\partial \omega}{\partial k}$, we obtain the velocity of the axoplasmic pressure pulse in the high viscosity limit: 
\begin{equation}
v_{gr}=c \cdot \alpha \cdot v_0 = c \cdot R \sqrt{\frac{\omega}{\mu(\kappa +\frac{2R}{K})}}.
\end{equation}
Under the same limit, the distance over which the pulse amplitude decreases $e$-fold (the decay length), is \cite{rayl45,morg54,rubi78}:
\begin{equation}
l = \frac{c}{2} \cdot R \sqrt{\frac{1}{\omega \mu (\kappa +\frac{2R}{K})}}.
\end{equation}

\subsection{Myelinated axons}

 In the H-H model of propagation of the action potential, the role of the myelin sheath is to insulate the membrane electrically and decrease its electrical capacitance; both these effects contribute to more rapid conduction of the action potential. However, the myelin sheath is also quite rigid. About 80\% of dry myelin content is lipid, with a roughly 2:2:1 molar ratio of the three major lipids, cholesterol, phospholipids and galactolipids \cite{knaa05}. Cholesterol is known to increase the strength and elastic modulus of lipid bilayers, with the highest modulus and highest strength achieved for 50 mol\% cholesterol \cite{need95}, i.e., rather close to the concentration in myelin. The remaining 20\% of dry myelin content consists of protein, including various types of fibrous proteins that can form rigid structures. Altogether, for the deformation of circumferential strain, the myelin elastic area expansion modulus is about 2 N/m for a single myelin layer of 4 nm thickness \cite{need95}. Noting that $K =Eh$ (for $\nu=0.5$), the myelin effective Young's modulus is $E = 5\cdot10^8$ Pa. The ratio of the axon diameter (internal myelin diameter) to the nerve fiber diameter (external myelin diameter) is typically about 0.7  \cite{dona05}, implying a myelin thickness of $h\approx0.43 R$. Therefore, $\frac{2R}{K}=\frac{2R}{Eh} \approx 9.3\cdot 10^{-9}$ Pa$^{-1}$. This is about 23 times greater than the compressibility of water at 38$^\circ$C, $\kappa =4.04\cdot10^{-10}$ Pa$^{-1}$. Hence, $v_0$, as used in Eq. (4), is only $\sqrt{24}\approx4.9$ times less than the speed of sound in water (see Eq. (1)). In other words, the myelin sheath is so rigid that its distensibility decreases the pulse speed only about 4.9 times compared to the theoretical maximum if axon walls were absolutely rigid. For typical myelinated axons, e.g., cat myelinated axons of 10 $\mu$m fiber diameter at 38 $^\circ$C, Eq. (4) yields $v_{gr} = 88$~m/s (using $E=5\cdot10^8$ Pa, $2R=7 \mu$m, $h=0.43 R$, $\nu=0.5$, $\omega=3142$ rad/s, $\mu=6.82 \cdot 10^{-4}$~Pa$\cdot$s and $\kappa =4.04\cdot10^{-10}$~Pa$^{-1}$). Given the roughness of the estimate, this theoretical value of 88 m/s is in excellent agreement with the measured action potential velocity of 60 m/s for these fibers \cite{hurs39}.
From Eq. (5), the decay length for the propagation of the pulses in myelinated segments of such a fiber (where the pulse propagates passively) is 14 mm. Assuming that the distance between the nodes of Ranvier is 1 mm (100 times the external diameter of myelin \cite{stru04}), the pulse amplitude decays 7\% over the internodal distance.

\subsection{Unmyelinated axons}

Unmyelinated nerve fibers usually have diameters of 0.1 - 1.2 $\mu$m \cite{hobb07}. The action potential conduction velocity, in m/s, is given approximately by $v\approx 1800\sqrt{R}$, where $R$ is the axon radius in meters \cite{hobb07} (values quoted in the literature range from $v\approx 1000\sqrt{R}$~\cite{plon88} to $v\approx 3000\sqrt{R}$ \cite{rush51}).
Therefore, for an axon of 1~$\mu$m diameter, conduction velocity is about 1.27~m/s.
Assuming $2R=1$ $\mu$m, $K=0.8$ N/m (twice the single plasma membrane value of 0.4 N/m \cite{thom97}), and $\omega$, $\mu$, $\kappa$ and $\nu$ as before for myelinated axons, Eq. (4) yields a pressure pulse velocity of $v_{gr} = 1.11$ m/s, very close to the experimental value. From Eq. (5), the decay length for this fiber is 0.18 mm. It is truly remarkable that, for both myelinated and unmyelinated fibers, the predicted velocity of propagation of axoplasmic pressure pulses is very close to the measured velocities of nerve impulses. 

\subsection{Dependence of propagation velocity on fiber diameter}

As can be seen from the above numerical estimates, for both myelinated and unmyelinated axons, the condition $\frac{2R}{K} \gg \kappa$ is true; and therefore, Eq. (4) simplifies to:
\begin{equation}
v_{gr}= \sqrt{\frac{2 R \omega K}{3 \mu}}.
\end{equation}
This relationship shows that for unmyelinated axons, the axoplasmic pressure pulse velocity is proportional to the half-power of the axon diameter, assuming that the pulse duration, the membrane area expansion modulus and the axoplasm viscosity do not vary with axon diameter. For myelinated axons, introducing $\gamma =h/R$ (the ratio of the myelin sheath thickness to the axon radius) and using $K=Eh$, we can rewrite Eq. (6) as:
\begin{equation}
v_{gr}= R\sqrt{\frac{2\omega E \gamma}{3 \mu}}.
\end{equation}
Therefore, for myelinated axons, assuming that pulse duration and specific properties of axoplasm and myelin are independent of diameter and that the myelin sheath thickness scales in proportion to the fiber diameter (i.e., $\gamma$ is constant), the pressure pulse velocity scales linearly with the fiber diameter. It is interesting that a linear dependence of pressure pulse velocity on fiber diameter is also obtained for myelinated axons without assuming $\kappa\ll\frac{2R}{K}$, from Eq. (4):
\begin{equation}
v_{gr}= c \cdot R \sqrt{\frac{\omega}{\mu(\kappa +\frac{2}{E \gamma})}}.
\end{equation}
In summary, given that specific properties of axoplasm and myelin are independent of fiber diameter and that the myelin sheath thickness scales in proportion to fiber diameter, the axoplasmic pressure pulse velocity scales as $\sqrt{R}$ for unmyelinated axons and as $R$ for myelinated axons, similar to what is measured experimentally \cite{hobb07,plon88,rush51}. 

\subsection{Dependence of propagation velocity on temperature}

The duration of the action potential increases with decreasing temperature, e.g., with Q$_{10}$ (the ratio of the duration at one temperature to the value at a temperature 10 $^\circ$C lower) of 3.4 at 37 $^\circ$C for cat vagus myelinated fibers \cite{paint66}. Assuming that the duration of the axoplasmic pressure pulse is similar to that of the action potential, taking the change in $\mu$ from 37 $^\circ$C to 27 $^\circ$C as for water, Q$_{10}=0.81$ \cite{lide02}, and assuming that the other parameters in Eq. (4) do not change with temperature, we obtain the predicted Q$_{10}$ of 2 for the nerve impulse propagation velocity. This is rather close to the experimental value of 1.6 \cite{paint65}.

\section{Meyer-Overton Rule and the Action of Anesthetics}

If axoplasmic pressure waves are involved in a meaningful way in a normal propagation of the nerve impulse, we arrive at a possible explanation of the Meyer-Overton rule for inhaled anesthetics \cite{meye99,over01,urba06}. Essentially, the rule states that anesthetic potency is determined by the bilayer concentration of anesthetic, independent of its molecular identity (e.g., Ref.~\cite{cant01}). Therefore, the Meyer-Overton rule can be explained if there is a property of the lipid bilayer that is essential for nerve impulse propagation and that, in addition, depends only on the bilayer concentration of anesthetic. As is shown below, one such property may be the bilayer area expansion modulus, which affects both the decay and the velocity of axoplasmic pressure pulses. It is also shown that anesthetics may inhibit the mechanical modulation of Nav channels by the traveling pressure pulse (and therefore contribute to suppression of conduction of the action potential) in a manner consistent with the Meyer-Overton rule.

Suppose that an anesthetic molecule is dissolved within the lipid bilayer (anywhere from the headgroup to the hydrophobic region), and it weakens and destabilizes the bilayer structure nonspecifically, e.g., through creating a defect in the packing of lipid molecules and thereby nonspecifically disordering lipid chains (Fig. 4(a)).
\begin{figure}[t]
\includegraphics[width=9cm]{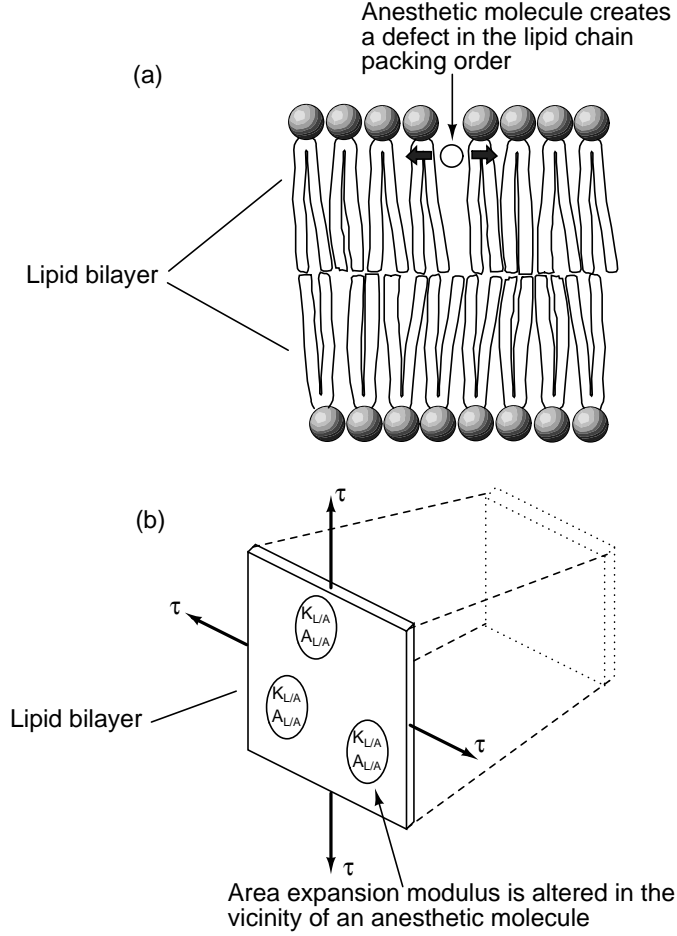}
\vspace*{8pt}
\caption{\label{fig5} (a) Disordering of lipid packing order by an anesthetic molecule dissolved in the bilayer. (b) Pockets of altered area expansion modulus in the vicinity of anesthetic molecules.}
\end{figure} 
Such disordering interaction will generally decrease the bilayer elastic area expansion modulus in the vicinity of the lipid/anesthetic molecular complex (it is well established that bilayers with larger lipid chain disordering generally display lower elastic moduli and tensile strengths \cite{need95}). Using a simple macroscopic model of the membrane \cite{need95}, shown in Fig. 4(b), the total membrane area expansion modulus can be expressed as: 
\begin{equation}
K=\left(\frac{a_M}{K_M}+\frac{a_{L/A}}{K_{L/A}}\right)^{-1},
\end{equation}
where the right-hand side of the equation is a combination of the area expansion moduli of the two components, namely, non-anesthetized membrane $K_M$ (which describes the elasticity of the lipid bilayer, embedded proteins and other native structures) and a purported lipid/anesthetic molecular complex $K_{L/A}$ ($K_{L/A}<K_M$), scaled by their area fractions $a_M$ and $a_{L/A}$. Assuming that anesthetic molecules are dissolved only in the lipid segments of the membrane and that they do not cluster together, the area fractions $a_M$ and $a_{L/A}$ are given by: 
\begin{equation}
a_{L/A} = 1 - a_M = n_{A} \cdot h \cdot A_{L/A},
\end{equation}
where $n_{A}$ represents the concentration of anesthetic molecules within the membrane, in units of molecules per volume, $h$ the bilayer thickness, and $A_{L/A}$ the membrane area occupied by a single lipid/anesthetic molecular complex. Note that $A_{L/A}$ and $K_{L/A}$ are independent of the chemical identity of the anesthetic, due to the purported nonspecificity of anesthetic/lipid interaction. Combining Eqs. (9) and (10):
\begin{equation}
K=\left(\frac{1- n_{A} h  A_{L/A}}{K_M}+\frac{ n_{A}  h  A_{L/A}}{K_{L/A}}\right)^{-1},
\end{equation}
where the right-hand side of the equation does not depend on the identity of the anesthetic, but only on its bilayer concentration $n_A$. As was shown above, both the velocity and the decay length of axoplasmic pressure pulses vary as $\sqrt{K}$. Therefore, a decrease in the area expansion modulus $K$ due to an increase in the anesthetic concentration $n_A$ will reduce the pressure pulse velocity and increase its decay, which, when $n_A$ is of sufficient magnitude, may lead to an inhibition of the pulse propagation. From the form of Eq. (11), it follows that anesthetics acting in such a way will comply with the Meyer-Overton rule. From the line of argument above, it is clear that any nonspecific anesthetic/lipid interaction that results in decreasing $K$ with increasing $n_A$ and thus leads to a block in nerve impulse conduction will be consistent with the Meyer-Overton rule.

Another possible mechanism of action of anesthetics consistent with the Meyer-Overton rule relies on inhibition of modulation of Nav channels by the traveling pressure pulse, in cases where such modulation is mediated by changes in the membrane lateral pressures (Fig. 3(a)). As before, we assume that anesthetic molecules dissolved in the bilayer weaken it nonspecifically and that increasing anesthetic concentration $n_A$ decreases the membrane area expansion modulus $K$. As was discussed above, the potential energy of a propagating axoplasmic pressure pulse is mainly stored in the membrane strain. The membrane strain energy, $E_p$, is proportional to $L\Delta\tau^2/K$, where $L$ is the pulse length and $\Delta\tau$ is the peak increase in the membrane tension $\tau$ from equilibrium. Since $L\sim\sqrt{K}$ (assuming a fixed pulse duration, see Eq. (6)), for the same pulse energy ($E_p$ fixed), it follows that $\Delta\tau\sim K^{1/4}$. Therefore, a decrease in the area expansion modulus $K$ will result in a reduced peak tension increase $\Delta\tau$. At sufficiently high $n_A$ (low $K$), the decreased $\Delta\tau$ may not suffice to stretch-modulate Nav channels to the degree that is required for the process of co-propagation of the electrical and mechanical excitations to occur, possibly leading to conduction block. It should also be noted that, if the anesthetic/lipid interaction changes the lipid bilayer area without affecting its area expansion modulus, the equilibrium tension in the membrane may change, which, in turn, should affect gating of Nav channels and also possibly lead to conduction block. This anesthetic/lipid interaction would also conform to the Meyer-Overton rule if it is nonspecific. In this context, it should be noted that an anesthetic/lipid interaction that alters a mechanical property of the lipid bilayer in a nonspecific manner would probably involve a defect-like structure, in which physical properties of the anesthetic compound are not important provided that the compound satisfies certain criteria for creating the defect (e.g., being hydrophobic, nonpolar, and nonreactive and possessing certain spatial and mass dimensions).

\section{Discussion}

It has recently been suggested \cite{heim05,heim07} that the H-H model \cite{hodg52} of the action potential does not provide a satisfactory description of the nerve impulse because it does not include the mechanical \cite{iwas80b,iwas80,tasa80,tasa82,tasa90} and optical \cite{tasa68,tasa69} changes associated with the action potential and because it is inconsistent with observed reversible changes in temperature and heat \cite{abbo58,howa68,ritc85,tasa89,tasa92}. Furthermore, the famous Meyer-Overton rule \cite{meye99,over01,urba06} is incongruous with protein models of action of anesthetics \cite{heim07}, which are consistent with the H-H model. A theory of the action potential based on mechanical solitons propagating in the axon membrane was suggested instead \cite{heim07}. Although this theory is in principle able to account for the above-mentioned phenomena, apparently it cannot explain changes in the propagation velocity that are related to axon diameter.

Following related earlier work \cite{rva03,rva09}, here we propose that the mechanical, optical and thermal changes associated with the action potential \cite{iwas80b,iwas80,tasa80,tasa82,tasa90,abbo58,howa68,ritc85,tasa89,tasa92,tasa68,tasa69} are the result of an axoplasmic pressure pulse that accompanies the action potential and has a roughly similar duration. We have shown that, for both myelinated and unmyelinated axons of different diameters, such pressure pulses propagate with a velocity close to the measured action potential velocity. Hence, any axoplasmic disturbance resulting from a propagating action potential (e.g., caused by the influx of extracellular Ca$^{2+}$ ions and their presumed action on acto-myosin cytoskeletal elements, or the disturbance caused by voltage-induced membrane movement \cite{zhan01}), should accumulate into a larger, shock-like axoplasmic pressure wave such as we propose. Further, based on known axon properties, we suggest a model in which the axoplasmic pressure wave accelerates propagation of the H-H type excitation wave, leading to co-propagation of the two waves at the velocity of the axoplasmic pressure wave. We also suggest several mechanisms for pressure wave amplification that could counteract its expected viscous decay. The presented model has experimental basis in the above-mentioned observations of mechanical, optical and thermal changes associated with the action potential, which all point to the involvement of a mechanical or thermodynamic process in the propagation of the nerve impulse. Importantly, the simple and powerful Meyer-Overton rule \cite{meye99,over01,urba06}, valid over five orders of magnitude, currently appears to be difficult to interpret in the context of presently accepted models. Within the presented framework, however, the Meyer-Overton rule can be explained based on known interaction of anesthetics with lipid bilayers. A simple test of the presented theory is its ability to predict the velocity of the nerve impulse. We have shown that, over two orders of magnitude (for myelinated and unmyelinated axons), the predicted velocity is close to that which is experimentally observed, despite the fact that the precision of calculations is limited by current knowledge of the mechanical properties of axons. The dependence of nerve impulse velocity on fiber diameter and temperature is also reproduced well by the model. 

If some types of neurons or other excitable cells indeed rely to a significant degree on the presented mechanism of signal transmission, perhaps so much so that propagation of the signal within the pure H-H framework is not feasible, it becomes plausible that various pressure-mediated processes may play important roles in the function of these cells. Depolarization of the axon hillock to the excitation level, which presumably leads to a radial contraction of the hillock membrane and the initiation of an axoplasmic pressure pulse, may proceed not only through a passive spread of ionic currents from the postsynaptic sites, but also through mechanical modulation of gating of ion channels in the vicinity of the axon hillock. Therefore, synaptic integration may include relay of cytoplasmic pressurization and displacement from postsynaptic sites through the dendritic tree to the axon hillock. It follows that learning and memory in neurons may be partially reflected in the strength and duration of a pressure-generating contraction of a protein filament network that is anchored to the postsynaptic membrane; the contraction may be activated as a result of the development of postsynaptic potentials. Hence, some of the structural elements that underlie learning and memory may be similar to those that regulate muscle cell strength (e.g., the number of actin-myosin filaments), which would provide a possible link between learning and memory in neurons and the changes that occur in muscle cell strength due to exercise. The existence of a dense undercoating just beneath the axon membrane in the axon hillock and nodes of Ranvier (e.g., Ref.~\cite{shep94}), currently functionally unexplained, may in fact be the putative contractile filament network attached to the inner membrane surface, expressed specifically in regions where energetic generation and amplification of the axoplasmic pressure pulses has to occur. Nerve impulses generated in the axon hillock may affect processes in the neuronal cell body, affecting dendrites and contacting synapses through back-propagating cytoplasmic pressure pulses. Action potential generation and propagation may be affected as a result of stretch-modulation of membrane voltage-gated channels by bilayer stresses resulting from altered [Ca$^{2+}$]$_\textnormal{i}$ levels and their action on cytoskeletal elements following previous spike activity. Pressure pulses associated with action potentials may play a role in transport of substances in excitable cells. Synaptic vesicle release at axon terminals may be facilitated by a contraction of the knob-shaped membrane skeletal protein network of the synapse, creating a force pushing the vesicles toward the synaptic cleft. The major components of the model, propagating hydraulic pulses and radially contracting tubes generating an axial force on the tube contents, are exploited in the cardiovascular and gastrointestinal systems, suggesting a possible evolutionary link between these systems and excitable cells. Finally, it is suggested that the evolutionary predecessor of the ability to transmit nerve impulses may be the well-known ability of freely moving cells to control membrane movement by forces of microfilaments located just beneath the plasma membrane (e.g., Ref.~\cite{shep94}), in response to external as well as internal mechanical stimuli.

\section*{Acknowledgments}
The author gratefully thanks Prof. Michael A. Rvachov and Prof. William Bertozzi for useful discussions. The author also thanks anonymous reviewers for their thorough and helpful reviews. This work was partially supported by US DOE grant DE-FC02-94ER40818.




\end{document}